# Unravelling Negative In-plane Stretchability of 2D MOF by Large Scale Machine Learning Potential Molecular Dynamics

Dong Fan,[1,]* Aydin Ozcan,[2] Pengbo Lyu,[3] Guillaume Maurin[1,]*

[1] *ICGM, Univ. Montpellier, CNRS, ENSCM, Montpellier, 34095, France*
[2] *Turkey Marmara Research Center, Materials Technologies, Gebze, Kocaeli, 41470, Turkey*
[3] *Hunan Provincial Key Laboratory of Thin Film Materials and Devices, School of Material Sciences and Engineering, Xiangtan University, Xiangtan, 411105, People's Republic of China*

*Corresponding author
Email: dong.fan@cnrs.fr (D.F); guillaume.maurin1@umontpellier.fr (G.M)

Two-dimensional (2D) metal-organic frameworks (MOFs) hold immense potential for various applications due to their distinctive intrinsic properties compared to their 3D analogues. Herein, we designed *in silico* a highly stable $NiF_2(pyrazine)_2$ 2D MOF with a two-periodic wine-rack architecture. Extensive first-principles calculations and Molecular Dynamics simulations based on a newly developed machine learning potential (MLP) revealed that this 2D MOF exhibits huge in-plane Poisson's ratio anisotropy. This results into an anomalous negative in-plane stretchability, as evidenced by an uncommon decrease of its in-plane area upon the application of uniaxial tensile strain that makes this 2D MOF particularly attractive for flexible wearable electronics and ultra-thin sensor applications. We further demonstrated that the derived MLP offers a unique opportunity to effectively anticipate the finite temperature mechanical properties of MOFs at large scale. As a proof-concept, MLP-based Molecular Dynamics simulations were successfully achieved on 2D $NiF_2(pyrazine)_2$ with a dimension of 28.2 × 28.2 $nm^2$ relevant to the length scale experimentally attainable for the fabrication of MOF film.

# Introduction

Metal-organic frameworks (MOFs) have attracted substantial interest from both academic and industrial perspectives owing to their unique features such as high porosity, tunable structures and versatile functionalities.[1–3] An almost infinite number of MOF architectures with different pore size/shape and a wide range of chemical functionality can be thus designed by assembling a rational selection of inorganic and organic secondary building units.[4–6] Highly flexible MOFs have emerged as an intriguing sub-class of porous hybrid porous materials that undergo reversible/irreversible changes in their crystalline structures and pore dimensions (ligand flip, gating, swelling, breathing) in response to external stimuli such as guest adsorption, temperature or mechanical pressure.[7–10] Some of these flexible three-dimensional (3D) MOFs have demonstrated unusual mechanical properties, *i.e.*, superhigh flexibility, negative gas adsorption, and negative linear compressibility (NLC) among others that promoted new applications of this family of materials in the fields of gas separation, sensors and micromechanics.[8,11,12] Typically, the wine-rack MIL-53 built up of infinite trans chains of corner-sharing $MO_4(OH)_2$ octahedra (M=$Al^{3+}$, $Cr^{3+}$,…) and other related flexible MOFs have been computationally and experimentally proved to exhibit significant anisotropy in their Young's modulus and/or shear modulus at the origin of their anomalous mechanical properties.[13–20]

More despite an exhaustive list of mechanical studies have been reported on 3D MOFs over the last 20 years, much less attention has been paid on two-dimensional (2D) MOFs. One of the main reasons is that 2D MOFs only make up a tiny portion of all known MOFs; this is especially true for 2D MOFs with the wine-rack pore geometry, of which only a very small number of instances have been reported with a perfect square lattice (**sql**)topology.[21,22] Compared to their 3D analogues, 2D materials generally exhibit higher strength and toughness due to the limited number of constitutive atomic layers.[23] As a result, they can undergo plastic deformation under high strain without fracturing, making them more flexible, highly resilient, and capable of withstanding high stresses without breaking, and therefore they can be adjustable into a variety of shapes.[24,25] Therefore, the design of 2D MOFs with a wine-rack topology is anticipated to lead to inherent flexibility and potentially unusual mechanical properties, making them promising candidates for a wide range of applications, including flexible electronics like wearable devices, sensors, and smart patches.[24–27]

In this context, a 2D $NiF_2(pyz)_2$ (where **pyz** stands for pyrazine) with 2D wine-rack topology was constructed *in-silico* and its flexible behaviour was systematically investigated at different length scales by combining high-precision first-principles calculations and large-scale Machine learning Potentials (MLP)-based Molecular Dynamics (MD) simulations. The MLP strategy have been shown to significantly enhance the efficiency of molecular simulations by using high-quality datasets from first principles calculations to predict accurately material properties while reducing computational costs dramatically.[28–30] However, developing high-quality MLPs for MOFs is still a great challenge, because the training time increases exponentially with the increase of system size and number of chemical species.[28] To date, only few MLPs for MOFs have been reported, exclusively for 3D MOFs, including UiO-66, MOF-5, ZIF-8, and MOF-74.[31–35] Herein, systematic phonon spectrum calculations and first-principles MD simulations first revealed that the proposed 2D MOF material does not exhibit any imaginary phonon modes in the whole Brillouin zone and it can maintain its structure integrity at room temperature as well as in aqueous solution. Large-scale modelling of the MOF system, *i.e.* length scale > $10 \times 10$ nm$^2$ with over thousands of atoms and time scale of 1 ns, largely beyond the realm of first-principles calculations, was further achieved by developing the first MLP for 2D MOF that was implemented into a MD scheme. This novel MLP was first validated by a fair reproduction of the first-principles calculations performed on 2D $NiF_2(pyz)_2$, *e.g.* energy profiles, phonon properties and thermodynamics data. Its implementation into a MD-scheme enabled to explore the structural responsiveness of the 2D MOF to external tension strain at finite temperatures. The MLP-based MD simulations revealed an enormous in-plane flexibility of the 2D $NiF_2(pyz)_2$ structure (originating from the notable difference in Poisson's ratio along the zigzag-direction [0.86] and armchair-direction [0.14]), with a resulting simulated fracture strain up to 19% even at room temperature, similar to the value reported previously for graphene.[36] Decisively, we revealed an anomalous negative in-plane stretchability of the 2D NiF$_2$(pyz)$_2$ upon uniaxial stretching both at 0 K and finite temperature. This intriguing phenomenon implies a counterintuitive decrease of the MOF in-plane area upon the application of a wide range of uniaxial stretching below the fracture strain. Such unconventional mechanical response may have potential applications in wearable devices and flexible electronic material applications. Beyond the design of a novel 2D MOF material with unconventional mechanical properties, we further applied the well-trained MLP to a large scale 2D NiF$_2$(pyz)$_2$ atomistic model (size ~28.2 × 28.2 nm$^2$ corresponding to 36,800 atoms) that reduces the gap between experimental and theoretical attainable length scales in the field of MOFs, of key importance to venture into phenomenon yet to be discovered.

## Results

**Structure and stability of the 2D NiF$_2$(pyz)$_2$ at the quantum level.** We first built a structure with metal node and **pyz** linker arranged in a 2D-array. Three initial atomistic models were considered corresponding to distinct orientations of **pyz** and Ni$^{2+}$ in ***ab*** planes with dihedral angles ($\angle_{N-Ni-N-C}$) of ± 45° and 90° leading to different space group symmetry as shown in Supplementary Fig. S1. These 3 models were geometry optimized at the Density functional theory (DFT) level and converged towards a perfect square planar geometry (tetragonal space group *P4/mmm*) as shown in Supplementary Fig. S2. The resultant 2D NiF$_2$(pyz)$_2$ geometry, depicted in Fig. 1a, consists of **pyz** moieties coordinated to two Ni atoms *via* their opposite N atoms. Each Ni atom is coordinated to four **pyz** moieties and two fluorine atoms, featuring a perfect 2D **sql** topology with ***a*** = ***b*** = 7.061 Å, ***α*** = ***β*** = ***γ*** = 90° as defined at the DFT+*U*[37] level (*cf.* Supplementary Table S1 for the lattice parameters simulated by using other functionals). The proposed 2D NiF$_2$(pyz)$_2$ structure exhibits a pattern resembling a wine-rack motif that has been only rarely reported so far for 2D MOFs[21,22,38,39].

To gain more structural details of the 2D NiF$_2$(pyz)$_2$, we employed the Tersoff and Hamann approximation[40] in constant height mode to simulate its scanning tunneling microscopy (STM) image, as depicted in Fig. 1b. This approach mimics the standard experimental setup applied to collect STM images of MOFs,[41] delivering a precise description of the structure at the atomic level. The simulated STM image of NiF$_2$(pyz)$_2$ reveals a symmetrical square grid shape, resembling the 3D **Nb-OFFIVE**-Ni MOF along [001] direction.[42] Typically, the theoretical X-ray diffraction (XRD) pattern alongside the simulated Raman spectrum of 2D NiF$_2$(pyz)$_2$ are provided in Fig. 1c, to assist the identification of such 2D MOF in future experimental studies.

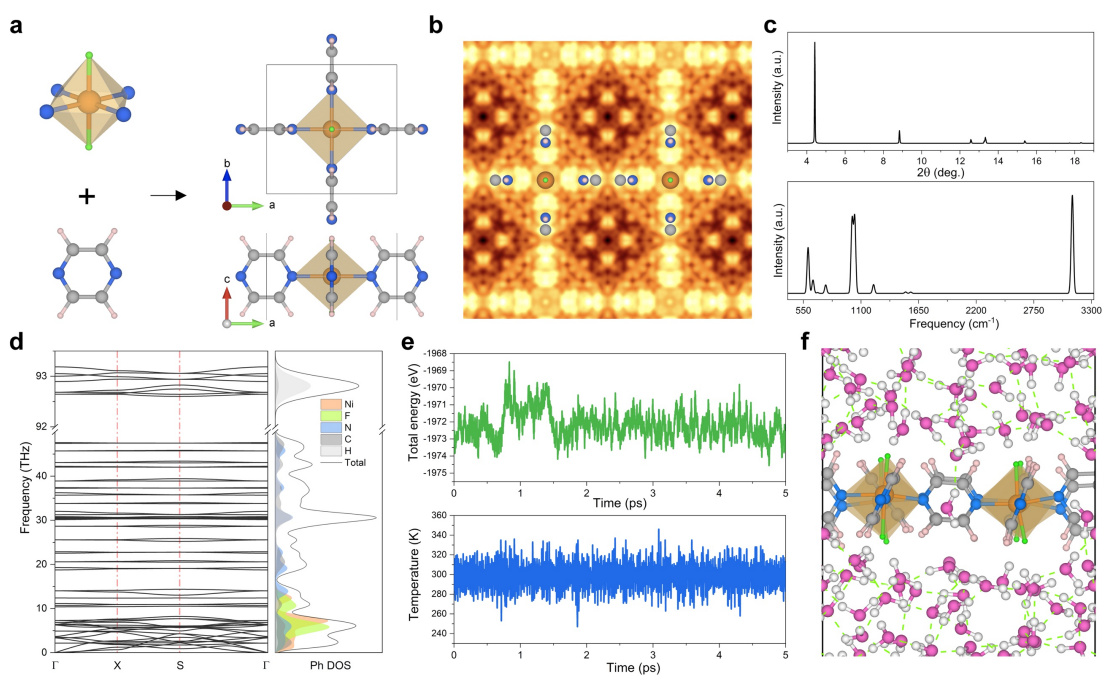

**Fig. 1 | Structure and stability of the 2D NiF$_2$(pyz)$_2$ explored at the quantum level.** (**a**) Side and top views of the DFT+$U$-geometry optimized structure of the 2D NiF$_2$(pyz)$_2$. Ni is shown in orange, N in blue, F in green, H in light pink, and C in dark-gray. The black line delimits the unit cell. (**b**) Simulated STM image of the DFT+$U$-geometry optimized 2D NiF$_2$(pyz)$_2$, by using the constant height mode at V$_{bias}$ = 1.5 V. The corresponding atoms are also superimposed on the STM image with the same color code than in (a). (**c**) Simulated XRD pattern (top) and Raman spectrum (bottom) of the 2D NiF$_2$(pyz)$_2$. (**d**) Simulated phonon spectrum and the corresponding phonon density of state (Ph DOS) for the 2D NiF$_2$(pyz)$_2$. (**e**) Evolution of the total energy (top) and temperature (bottom)

obtained during the AIMD simulations for 2D NiF$_2$(pyz)$_2$ (2 × 2 supercell) in aqueous solution. AIMD simulations conducted for 5 ps in the *NVT* ensemble (T=300 K) and the consideration of a timestep of 0.5 fs. (**f**) Side view of a representative snapshot of the 2D NiF$_2$(pyz)$_2$ in aqueous solution at 300 K after 5 ps-AIMD simulations. The color code of the atoms is the same as described before. Water molecules in the snapshot are marked by magenta (O) and white (H). The green dashed line represents hydrogen bonding interactions between water molecules.

The dynamic and thermal stabilities of the resulting 2D NiF$_2$(pyz)$_2$ are key features prior envisaging further applications. To evaluate these stabilities, high-precision calculations of the phonon spectrum were performed based on *ab initio* molecular dynamics (AIMD) simulations at 300K. Notably, there is no observable imaginary frequency present in the phonon dispersion curves as shown in Fig. 1d, suggesting that the 2D NiF$_2$(pyz)$_2$ is dynamically stable. The peak associated with the highest frequency around 94.5 THz (~3118 cm$^{-1}$), corresponds to the vibrational stretching mode of the C-H bond (Supplementary Movies S1-S3), while the frequency of metal-node is mainly localized in the low frequency range (below 15 THz). AIMD simulations were further performed at 300 K for 10 picoseconds (ps) to further assess the thermal stability of the 2D NiF$_2$(pyz)$_2$ at ambient conditions. The resulting energy fluctuation of the system over time, as shown in Supplementary Fig. S3, is very limited, suggesting that the 2D NiF$_2$(pyz)$_2$ structure remains intact without undergoing reconstruction during the simulation period. Therefore, 2D NiF$_2$(pyz)$_2$ exhibits both dynamic and thermal stabilities. Another important requirement for potential applications is the hydrolytic stability of the designed 2D NiF$_2$(pyz)$_2$. While many MOFs may not have long-term water stability,[43] it is critical to investigate the structural integrity of the 2D NiF$_2$(pyz)$_2$ in an aqueous media using AIMD simulations. As shown in Figs. 1e-1f, the resulting simulations performed in water at 300 K demonstrated that both temperature and total energy remain constant during the AIMD calculations and analysis of the 2D NiF$_2$(pyz)$_2$ frameworks over the overall AIMD simulations evidences the absence of structural change, supporting a high robustness of the 2D NiF$_2$(pyz)$_2$ (Supplementary Notes and Fig. S4 for the details).

**MLP training and validation for the 2D NiF$_2$(pyz)$_2$.** The target MLP was trained on the energy and force determined at the DFT level for the 2D NiF$_2$(pyz)$_2$. We applied a variety of strategies, *e.g.* static calculations with different supercells, AIMD simulations at different temperatures, and datasets under the application of different stresses to expand the diversity of the structure space (Supplementary Fig. S5). To further validate the accuracy of the trained MLP, the energies and forces were calculated for a test set of 1000 structures initially generated by AIMD at 300 K, and compared with the values obtained at the DFT level. The accuracy of the trained MLP is demonstrated by the nearly linear fit of the predicted MLP-derived energies (forces) versus the DFT-calculated energies (forces), as shown in Figs. 2a-2b, respectively. The root mean square errors for the MLP-derived energy and force are 0.083 meV atom$^{-1}$ and 0.013 eV Å$^{-1}$, respectively. This result reveals a remarkable accuracy of the developed MLP for capturing the properties of the trained systems. The strain-energy relationship of the 2D NiF$_2$(pyz)$_2$ was calculated for further validation of the trained MLP, as the accurate description of the energies under strain conduction is essential to avoid unphysical phase dissociation/reconstruction during the simulations. As depicted in Fig. 2c, our findings proved that the trained MLP describes very well the energetics of the strained phases referring to DFT data, suggesting that the trained MLP can be effectively applied to configurations not included in the training set. Fig. 2d shows that the phonon spectrum predicted by the MLP approach is in good accordance with the reference data obtained at the DFT level using density functional perturbation theory method.[44] The phonon density of states (Ph DOS) is also predicted accurately with peak positions well reproduced. (Fig. 2e). This demonstrates that our MLP model predicts well not only pair potentials and forces but also collective phonon dynamics of the lattice. Thermal properties, such as the Helmholtz free energy and the specific heat capacity ($C_V$), computed using the phonon relationships reproduced also very well the corresponding DFT-data, as shown in Fig. 2f. Notably, our MLP was trained with AIMD simulations performed in the temperature range from 100~300 K, but we showed that it can still accurately predict thermal properties beyond room temperature, demonstrating its ability to anticipate properties over a wide range of temperatures accurately.

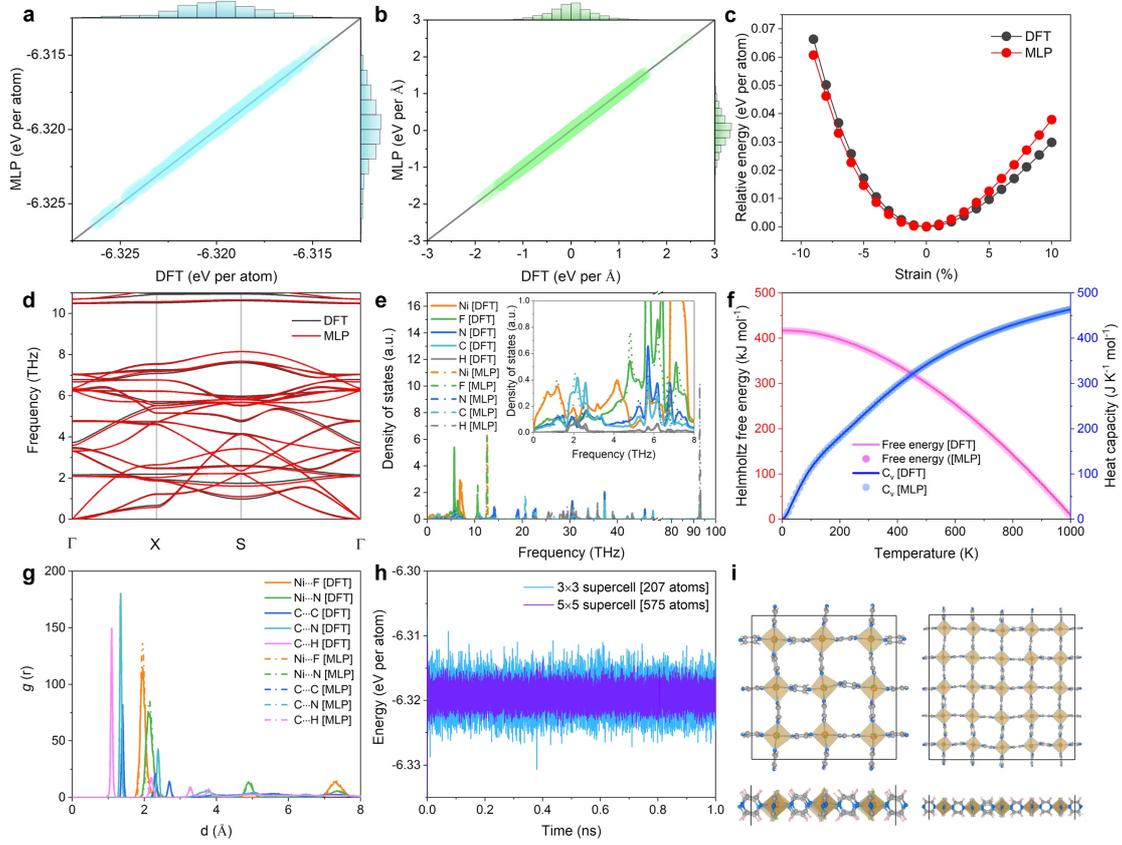

**Fig. 2 | Training, Validation and transferability of the MLP for the 2D NiF$_2$(pyz)$_2$.** MLP-derived (**a**) energies and (**b**) forces vs the corresponding DFT-values. (**c**) Strain-energy curves for 2D NiF$_2$(pyz)$_2$ simulated using MLP and DFT calculations. (**d**) Phonon dispersion of 2D NiF$_2$(pyz)$_2$ computed using MLP and DFT calculations (for clarity only dispersion curves below 11 THz are displayed). (**e**) Phonon density of states (Ph DOS) computed by using the DFT and MLP calculations, where the inset highlights the at the low frequency region. (**f**) Helmholtz free energy and specific heat capacity (C$_v$) computed using the DFT and MLP calculations as a function of temperature. (**g**) Radial Distribution Functions (RDFs) calculated for different atom pairs of 2D NiF$_2$(pyz)$_2$ at 300 K computed using MLP and AIMD, respectively. (**h**) Total energy fluctuations of the 2D NiF$_2$(pyz)$_2$ during the MLP-MD simulations at 300 K with 3 × 3 and 5 × 5 supercells. (**i**) Snapshots of the MLP-MD-derived NiF$_2$(pyz)$_2$ with 3 × 3, (left) and 5 × 5 (right) supercells

containing 207 and 575 atoms respectively after 1 ns MD simulation. Ni, F, N, C, and H atoms are marked with orange, green, blue, dark-gray, light-pink, respectively.

We further evaluated the capability of the trained MLP to describe the lattice dynamics of structure at finite temperatures (from 100~300 K). AIMD and MLP-MD simulations using the Nose–Hoover thermostat[45] were thus performed for 10 ps and 1.0 ns, respectively. The excellent agreement between the radial distribution functions (RDFs) calculated for different atom pairs of 2D $NiF_2(pyz)_2$ from both simulations at 300 K (see Fig. 2g) revealed the accuracy of our MLP in describing the structural behavior of 2D $NiF_2(pyz)_2$ at finite temperature. It should be emphasized here that AIMD simulations for MOFs are computationally expensive and usually run for only a few ps.[46] However, as evidenced by Figs. 2g-2h, based on a well-trained MLP, we can scale up the AIMD simulations to -ns scale, while maintaining an accurate description of the structure. We demonstrated that MLP trained using a relatively small lattice for $NiF_2(pyz)_2$ (*i.e.*, 3 × 3 supercell with 207 atoms) is transferable to an accurate exploration of the structural properties of a larger system (*i.e.*, 5 × 5 with 575 atoms) as shown in Figs. 2h-2i. Therefore, given the high accuracy and good transferability of our MLP in capturing interatomic forces of the 2D $NiF_2(pyz)_2$ overall system, one can expect to predict accurately the physical properties of this 2D MOF derived from the calculated force constants.

**Anisotropic in-plane mechanical properties of 2D $NiF_2(pyz)_2$.** Fig. 3a predicts a highly anisotropic in-plane flexibility of the 2D $NiF_2(pyz)_2$ once applying uniaxial tensile strain, which is unique for such 2D-wine-rack topology. To gain insight into this intriguing behaviour, the ideal tensile strength and critical strain were further explored by applying uniaxial strain along the zigzag and armchair directions, as shown in Fig. 3b. The uniaxial strain is defined as $\varepsilon = a - a_0/a_0$, where $a_0$ and $a$ are the lattice parameters of the original and strained 2D $NiF_2(pyz)_2$, respectively. For each applied strain, the lattice is also relaxed along the direction normal to the layer during the MLP-geometry optimization to ensure that the forces are minimal. The resulting strain-stress relationship is shown in Fig. 3c. The 2D NiF$_2$(pyz)$_2$ exhibits ideal strengths of 13 N m$^{-1}$ and 23 N m$^{-1}$ in the armchair and zigzag directions, respectively, with critical strains of 12% (armchair) and 40% (zigzag). Evidently, the critical strain along the zigzag direction is three times higher than along the armchair direction, indicating significant in-plane anisotropy of the mechanical property of 2D NiF$_2$(pyz)$_2$.

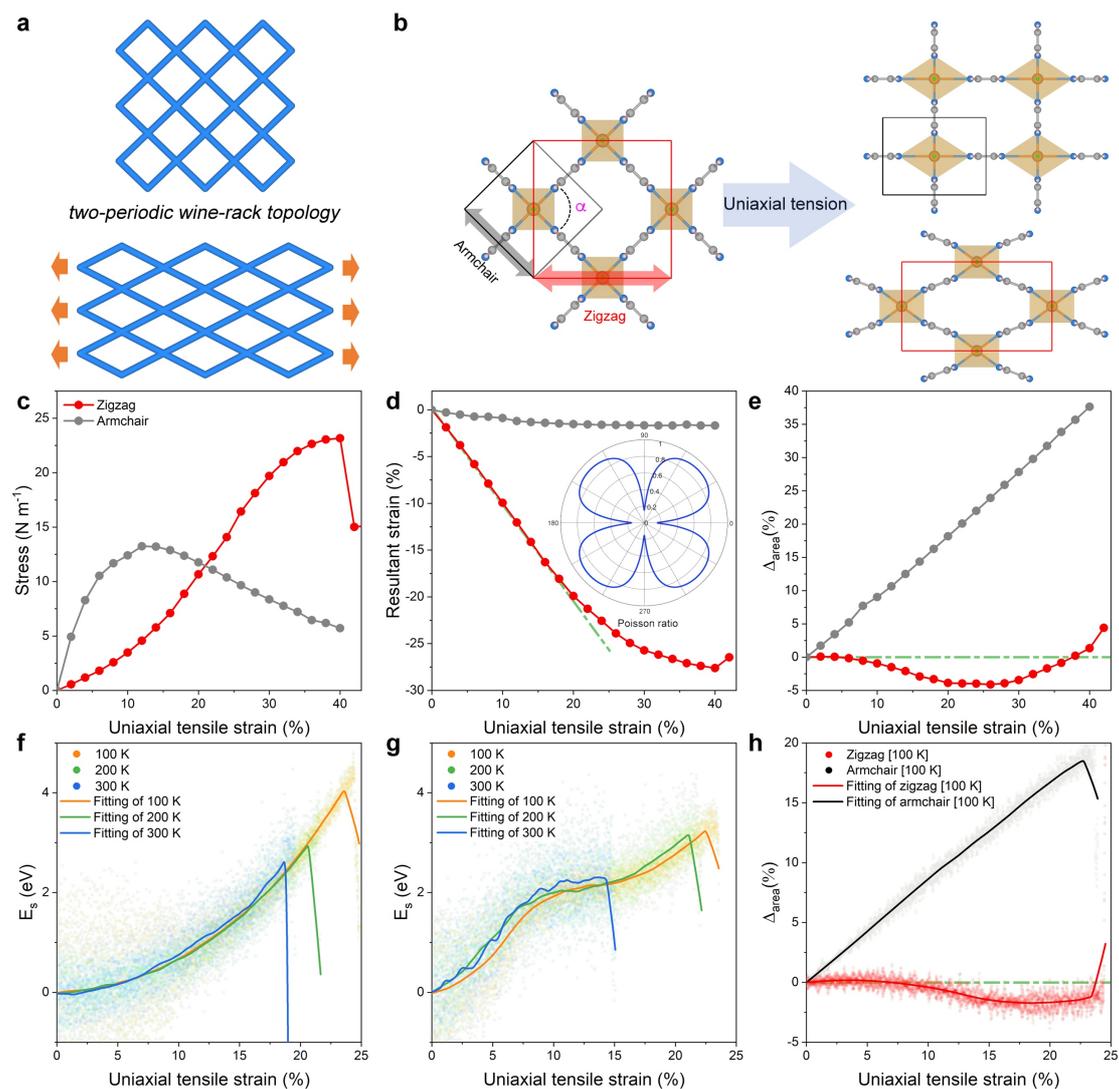

**Fig. 3 | DFT and MLP predicted structural response of 2D NiF$_2$(pyz)$_2$ under uniaxial in-plane tensile strain.** (**a**) Illustration of the deformation of a two-periodic wine-rack structure upon the uniaxial tensile strain. The blue stick is analogous to the ligand in the 2D NiF$_2$(pyz)$_2$, while the intersection of the hinges is the location of the metal nodes. (**b**) Schematic representation of the 2D NiF$_2$(pyz)$_2$ with uniaxial strain applied along different directions (zigzag and armchair) of the plane.

Color code: Ni: orange; N: blue; C: dark-gray; F: green; H: light pink. (**c**) Stress-strain responses of 2D NiF$_2$(pyz)$_2$ upon the armchair and zigzag strain. (**d**) Resultant strain induced by the uniaxial strain along zigzag and armchair directions. The inset shows the polar diagrams for Poisson's ratio of the 2D NiF$_2$(pyz)$_2$. (**e**) Relative in-plane area changes ($\Delta_{area}$, %) under a strain deformation along the two different directions. MLP-MD derived strain-energy ($E_s$) curves along the (**f**) zigzag and (**g**) armchair direction upon uniaxial tensile strain at different temperatures (100, 200, and 300 K). $E_s$ is defined as $E_s = E_{strain} - E_0$, where $E_{strain}$ is the total energy of the strained system and $E_0$ is the total energy of the original system. (**h**) MLP-MD derived $\Delta_{area}$ upon a strain deformation applied along the two different directions at 100 K. The colored dots represent the real MLP-MD simulation data points, and the solid lines represent the corresponding fitted data.

Fig. 3d and Supplementary Fig. S6 show that the resultant strain (with Poisson's ratio) and shear modulus along the zigzag direction are ~6 and ~20 times higher than along the armchair direction respectively, in line with a very large in-plane anisotropy of the mechanical properties of 2D NiF$_2$(pyz)$_2$. Such anisotropy is a signature of a substantial difference in terms of flexibility of the NiF$_2$(pyz)$_2$ along zigzag and armchair directions. It should be noted that the Poisson's ratios calculated from the linear fits of the resultant strain curve are consistent with that calculated using the elastic constants method, as shown in Supplementary Fig. S7. The resulting difference of 0.72 (0.14 along armchair and 0.86 along zigzag) is considerably larger than those reported for other systems including Tetraoxa[8]circulene-based COFs (0.04~0.384),[47] Fe$_2$(TCNQ)$_2$ MOF (0.41),[48] and M$_3$(C$_6$X$_6$)$_2$ (M = Co, Cr, Cu, Fe, Mn, Ni, Pd, Rh and X = O, S, Se) MOF (0~0.04),[49] that highlight an ultra-high in-plane flexibility of 2D NiF$_2$(pyz)$_2$. Such ultra-high in-plane Poisson's ratio combined with a high anisotropy makes 2D NiF$_2$(pyz)$_2$ a good candidate for practical large-magnitude strain engineering.[50]

**Negative in-plane stretchability of 2D NiF$_2$(pyz)$_2$.** Typically, when an in-plane tension strain (below the fracture strain) is applied to a 2D crystal, it will elongate in the direction of the applied tension and shrink on the other side, while resulting in an increase in change of the in plane-area ($\Delta_{area}$). This conclusion remains valid even for some materials with a negative Poisson's ratio, called auxetic materials.[51,52] In this respect, 2D NiF$_2$(pyz)$_2$ exhibits a standard behaviour with a DFT-predicted increase of $\Delta_{area}$ when a stretching is applied along the armchair direction (Fig. 3e). However, the situation is reversed when the stretching is applied along the zigzag direction with 2D NiF$_2$(pyz)$_2$ predicted to show an anomalous decrease of $\Delta_{area}$ as shown in Fig. 3e. In this later case, even at >30% strain, $\Delta_{area}$ still decreases. This unexpected phenomenon originates from the very high anisotropic in-plane flexibility of this 2D wine-rack MOF featuring high Poisson's ratio and high shear modulus along the zigzag direction (*cf.* Supplementary Fig. S7). As depicted in Supplementary Fig. S8, when the material is stretched, *i.e.* 24% along the direction of uniaxial tensile strain applied, the lattice parameters become significantly shorter on the opposite side along the zigzag direction, *i.e.* a decrease of 20% and 1.8% along the zigzag and armchair directions, respectively, leading to a decrease of $\Delta_{area}$. We further revealed that this variation is primarily due to structural changes occurring at the connections between the ligand and the metal node (Ni–N bonds namely), while the strain has almost no impact on the **pyz** ligand both in the armchair and zigzag directions (*cf.* Supplementary Figs. S9-S10). These conclusions were drawn from DFT-calculations performed at 0 K, while experimental nanoindentation measurements are usually performed at finite temperatures. However, incorporating the temperature effect into DFT calculations to investigate structural changes of materials under strain is theoretically highly challenging. Interestingly, we revealed above that our MLP-MD simulations performed at finite temperatures were also able to capture the same trend that that obtained by DFT at 0 K (Supplementary Movies S5-S6). Figs. 3f-3g, show that in the temperature range 100-300 K, MLP-derived energy of the structure increases more rapidly along the armchair direction than along the zigzag direction, as the applied uniaxial tensile strain increases, in line with the DFT calculations. In addition, we found that even at 300K, 2D NiF$_2$(pyz)$_2$ still has a high fracture strain of 19% (14%) along the zigzag (armchair) directions, indicating its exceptional flexibility under finite temperature conditions (see comparison with other 2D materials in Supplementary Fig. S11). We also observed the negative in-plane stretchability phenomenon in the $\Delta_{area}$ calculation at 100 K, as shown in Fig. 3h. At the same time, such phenomenon gradually becomes imperceptible as the temperature rises, as shown in Supplementary Fig. S12. This is because free-energy landscape of transitions typically gets smoother with increasing temperature and the probability of being trapped in local minima decreases. However, the material still exhibits a noticeably different response to tension changes along armchair and zigzag directions, even at 300 K. Therefore, both DFT and MLP-MD simulations demonstrated the unique negative in-plane stretchability phenomenon exhibited by the 2D $NiF_2(pyz)_2$ structure.

**Large-scale MLP applied to extended 2D $NiF_2(pyz)_2$ MOF models.** Our MLP strategy was applied above to a 2D $NiF_2(pyz)_2$ model with moderate size (up to 5 × 5 supercell with 575 atoms, ~ 3.5 × 3.5 $nm^2$), however, the dimensions of experimentally prepared 2D films are generally above 10 × 10 $nm^2$. Therefore, we extended the use of our MLP to 2D $NiF_2(pyz)_2$ models of much larger size, from 7.1 × 7.1 (2300 atoms) to 28.2 × 28.2 (36800 atoms) $nm^2$, as described in Figs. 4a-4f. It is clear that no distinct energy jump was detected in the energy fluctuations of the 2D MOF (Figs. 4a-4c) whatever the dimension of the system considered while the out-of-plane fluctuations of the structure tends to become apparent as the system size increases (*cf*. Figs. 4d-4f). We also evidenced that the 2D MOF containing >10 000 atoms, reach its equilibrium state only after ~20 ps of MLP MD simulations, which is almost unattainable in AIMD calculations with this size of system. Figs. 4g-4h show the RDFs calculated for different atom pairs of 2D $NiF_2(pyz)_2$ MOF models from the MLP-MD simulations performed on the 28.2 × 28.2 $nm^2$ model that compare very well with the AIMD-calculated profile on the small 2.0 × 2.0 $nm^2$ model. The MLP-MD simulated angular distribution function (ADF) reported in Fig. 4i also shows a single peak at ~90°, consistent with the angles formed by **pzy** ligand and Ni metal in the pristine 2D $NiF_2(pyz)_2$ configuration. The atoms are more mobile at higher temperature, resulting in a slightly broader features in the RDF and ADF at 300 K compared to that calculated at 100 and 200 K; but there is no quantitative change between these different temperature settings, and the MLP-based simulations reproduce all features obtained by AIMD on the small system.

All these results demonstrated the effectiveness and accuracy of our derived MLP in handling large 2D NiF$_2$(pyz)$_2$ system.

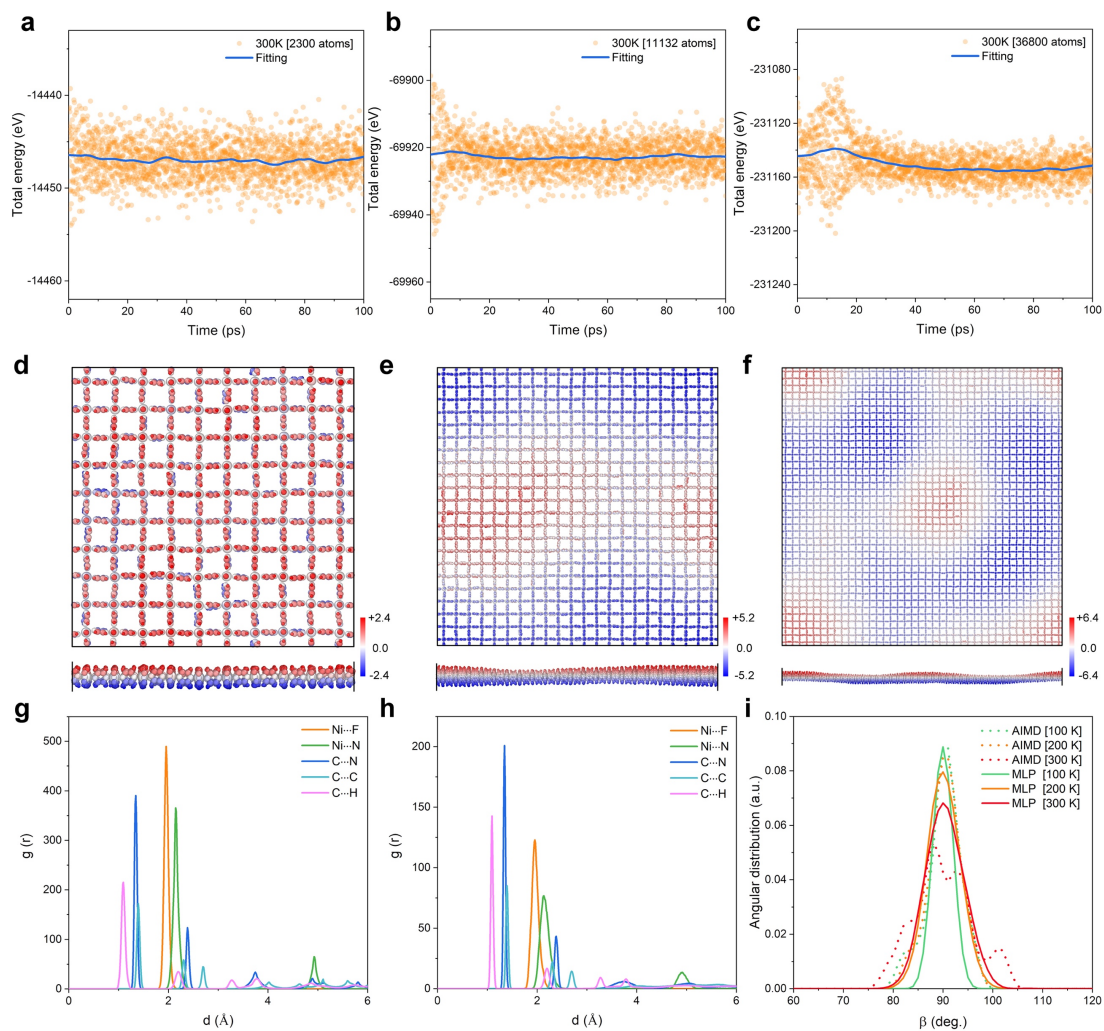

**Fig. 4 | Large scale MLP-MD simulations on extended 2D NiF$_2$(pzy)$_2$ models.** MLP-MD simulated fluctuation of the total energy of the 2D NiF$_2$(pzy)$_2$ with different sizes and a simulation time of 100 ps at 300 K, (**a**) 7.1 × 7.1 nm$^2$, (**b**) 15.5 × 15.5 nm$^2$, and (**c**) 28.2 × 28.2 nm$^2$, respectively. (**d-f**) The corresponding color coding represents the out-of-plane displacement at each system. The black boxes show the boundary of the periodic cell. Radial Distribution Functions (RDF) for different atom pairs of the 2D NiF$_2$(pzy)$_2$ at 300 K computed *via* (**g**) MLP-MD (28.2 × 28.2 nm$^2$) and (**h**) AIMD (2.0 × 2.0 nm$^2$) simulations in the *NVT* ensemble. (**i**) ADF ($\angle\alpha_{N-Ni-N}$) of 2D NiF$_2$(pzy)$_2$ at different temperatures from MLP-MD and AIMD simulations on the same 2D NiF$_2$(pzy)$_2$ systems as described in (**g-h**). The slightly more "jagged" appearance of the AIMD data

in panel **i** is due to the smaller number of configurations that are sampled from the MD trajectory compared to MLP data.

**Discussions**

We proposed and in-depth explored a super-flexible 2D MOF with wine-track architecture, 2D NiF$_2$(pyz)$_2$, by combining systematic first-principles calculations and MLP-large-scale MD simulations. The mechanical, thermal and water stability of 2D NiF$_2$(pyz)$_2$ was first evidenced through systematic phonon spectrum AIMD calculations. Interestingly, benefitting from the unique 2D wine rack motif, the 2D NiF$_2$(pyz)$_2$ was shown to exhibit an anisotropic in-plane flexibility with a fracture strain of up to 40% (0 K) and 17% even at room temperature, suggesting that this material can be used for wearable devices and flexible sensors. We also identified a counterintuitive negative in-plane stretchability phenomenon in 2D NiF$_2$(pyz)$_2$, whereby the surface area decreases with the applied tensile strain. Our systematic theoretical investigation showed that this behaviour originates from its unique 2D **sql** topology and anisotropic Poisson's ratio. Finally, we demonstrated that well-trained MLP can yield nanosecond MD simulations while maintaining first-principles accuracy, and can scale the MD simulation to more than tens of thousands of atoms. We believe that such a multi-scale computational approach presented here can greatly accelerate the theoretical design and understanding of novel flexible 2D MOF materials.

## Computational methods

**First-principle calculations.** The Vienna *ab initio* simulation package[53] (VASP, version: 5.4.4) was used to perform structural relaxation and total energy calculations with an energy cut-off of 900 eV (see more information regarding convergence testing in the *supplementary information*). Exchange-correlation potentials were treated within the generalized gradient approximation (GGA) employing the Perdew−Burke−Ernzerhof (PBE) functional.[54] The density functional theory (DFT) calculations used a 6 × 6 × 1 Γ-centered Monkhorst−Pack *k*-point sampling.[55] The DFT+*U* ($U$ = 6.4 eV) scheme[37] were used to account for the strong correlation of an unfilled d orbital of the Ni atom. The spin polarization was fully considered for all calculation process unless specifically mentioned. A 300 K AIMD simulation was run in the *NVT* (canonical ensemble) employing the Nosé-Hoover thermostat[45] to check the stability of 2D NiF$_2$(pyz)$_2$ structure at room temperature.

**Dataset preparations.** To obtain training data for all the MLP presented in this paper, Finite temperature *ab initio* molecular dynamics (AIMD) simulations based on DFT were performed using the VASP.[53] The PAW method and the GGA of the PBE type for the exchange-correlation functional was used during the AIMD simulations. The simulations consisted of more than 10,000 time-steps with a time step of 0.5 fs in 3 × 3 supercell. Brillouin zone sampling was performed using a Monkhorst-Pack *k*-point grid size of 1 × 1 × 1. All AIMD simulations were conducted in the NVT (canonical ensemble) employing the Nosé-Hoover thermostat[45] with a fixed temperature of 100 K, 200 K and 300 K.

**MLP development.** A deep neural network based approach was utilized to develop a MLP for the proposed NiF$_2$(pyz)$_2$ by using the DeePMD-kit package (version 2.0.1).[56] A total of 14788 snapshots were collected, with 13,788 used for MLP training and 1,000 used for MLP validation. The DeepPot-SE model was used to train the MLP by the DeePMD-kit code.[56,57] The model contains two networks: the embedding network and the fitting network. Both networks utilize the ResNet architecture.[58] The size of the embedding network was set to {25, 50, 100}. The size of the fitting network was set to

{240, 240, 240}. The cutoff radius was set as 6.0 Å and the smoothing parameter was set to 4.9 Å. It should be noted that the choice of cutoff in the MLP training is important and significantly affects the speed of training and the accuracy of the trained MLP, and we point out here that our choice of cutoff of 6 Å is sufficient for our 2D system to contain sufficient many-body information between adjacent atoms.[59,60] Throughout the training, we employed three layers of NN, where each layer consisted of 240 neurons. The loss is defined by:

$$L(p_\epsilon, p_f, p_\eta) = p_\epsilon \Delta\epsilon^2 + \frac{p_f}{3N}\sum_i |\Delta F_i|^2 + \frac{p_\eta}{9}||\Delta\eta||^2$$

Where $\Delta$ means the difference between the MLP results and the training data during the training process; $\epsilon$ is the energy per atoms; $N$ is the total number of atoms; $F_i$ is the force of atom i, and $\eta$ is the tensor divided by $N$. The number of batch and step of decay rate were set to 1,000,000 and 5,000, respectively. The initial learning rate was set as 0.001 and decays every 5,000 steps. The number of batches was set to 1,000,000.

**MLP-MD Simulations.** The MLP obtained from training process was utilized for MLP-MD simulations using the DeepMD-kit interface with the LAMMPS code.[56,61] To compute energy and force during MD simulations, the trained model was used as a pair style in LAMMPS.[61] We performed 1 ns long-time simulations in two systems, one system consistent with the AIMD simulation and one larger model initially, namely 3 × 3 and 5 × 5 supercells. To examine the applicability of our trained MLP, we conducted the large-scale molecular dynamics simulations for three large systems (including 2300, 11132, and 36800 atoms, respectively) of 100 ps length. MLP-MD simulations were executed within the NVT ensemble with Nosé-Hoover thermostat[45] at 300 K. The time step was set to 0.5 fs, and the periodic boundary conditions were employed along all three dimensions in the simulations. Two python-based tools, Phonopy and phonoLAMMPS,[62,63] were utilized to compute phonons for the 2D $NiF_2(pyz)_2$. The phonon bands for a 2 × 2 supercell were calculated with a consistent Brillouin-zone phonon sampling of 22 × 22 mesh. The path inside the Brillouin zone was from Γ (0.0, 0.0, 0.0) to X (0.5, 0.0, 0.0) to S (0.5, 0.5, 0.0) to Γ (0.0, 0.0, 0.0). VESTA and OVITO packages[64,65] were employed to plot the atomistic results.

## Data availability

All data needed to evaluate the conclusions in the paper are present in the paper and/or the Supplementary Information. The calculation data related to this article can be accessed online at https://www.github.com/xxxx. (will be available once the article is accepted)

## Code availability

The main codes used in this paper are VASP (https://www.vasp.at),[53] DeePMD-kit (https://github.com/deepmodeling/deepmd-kit),[56] and Lammps (https://www.lammps.org).[61] Detailed information related to the license and user guide are available at the referred papers and their websites.

## Acknowledgments

The computational work was performed using HPC resources from GENCI-CINES (Grant A0140907613).

## Author contributions

D.F and G.M. designed the research. D.F. carried out all the simulations with the help of A.O. and P.L. D.F., A.O. P.L., and G.M. wrote the manuscript. G.M. supervised the research.

## Competing interests

The authors declare that they have no competing interests.

## Additional information

Supplementary information

The online version contains supplementary material available at xxxx